\documentclass[a4paper]{jpconf}
\usepackage{graphicx}
\usepackage{amssymb}
\begin{document}

\title{The Luminosity Function for White Dwarfs in the SuperCOSMOS Sky Survey.}

\author{N. R. Rowell$^{1\bigstar}$, N. C. Hambly$^1$, P. Bergeron$^2$}

\address{$^1$ Institute for Astronomy, University of Edinburgh, Blackford Hill, Edinburgh EH9 3HJ\\
$^2$ D\'epartement de Physique, Universit\'e de Montr\'eal, C.P.~6128, Succ.~Centre-Ville,
Montr\'eal, Qu\'ebec H3C 3J7, Canada}

\ead{$^\bigstar$ nr@roe.ac.uk\\
\hspace{1.25cm}$^2$ bergeron@astro.umontreal.ca}

\begin{abstract}
We present the techniques and early results of our program to measure the luminosity function for White Dwarfs in the
SuperCOSMOS Sky Survey. Our survey covers over three quarters of the sky to a mean depth of $I\sim19.2$, and
finds $\sim9,500$ Galactic disk WD candidates on applying a conservative lower tangential velocity cut of $30$kms$^{-1}$. Novel techniques 
introduced in this survey include allowing the lower proper motion limit to vary according to apparent magnitude, 
fully exploiting the accuracy of proper motion measurements to increase the sample size. Our luminosity function
shows good agreement with that measured in similar works. We find a pronounced
drop in the local number density of WDs at a $M_{bol}\sim15.75$, and an inflexion in the luminosity function
at $M_{bol}\sim12$.
\end{abstract}

\section{Introduction}

The potential use of White Dwarfs (WDs) as chronometers has long been recognised. The relative accuracy of stellar age 
estimates based on up-to-date models, combined with the long evolutionary timescales of WDs and their abundance in most 
Galactic systems, means this approach has been used to age a wide variety of stellar configurations.
These include WD - subdwarf binaries (Monteiro et al. 2006), single burst populations such as globular and open 
clusters (Hansen et al. 2007; von Hippel \& Gilmore 2000, Jeffery et al. 2007), and continuous populations
such as the Galactic disk (Winget et al. 1987, Oswalt et al. 1995, Leggett et al. 1998).
%In the latter case, improvements made over several decades to the observational resources and stellar models
%available have refined age estimates to $8 \pm 1.5$ Gyr.
The standard tool used for this technique is the luminosity function (LF) for WDs, which contains information on both 
the absolute age and the star formation history of a stellar system. The star formation history is essentially
imprinted on the entire LF, and authors working on parametric forms for the LF (eg. Noh \& Scalo 1990) predict bumps and 
inflexions corresponding to changes in the rate of star production in the past. The absolute age is constrained by
the marked downturn in the LF at faint magnitudes, where the cooling ages of WDs converge on that of the population 
as a whole.

Modern surveys that apply this to the Galactic disk have suffered from small observational samples of faint WDs at the 
LF turning point. This is confounded by the fact that the atmospheric physics of such `ultracool' WDs is currently
poorly understood, with only a handful of stars extensively studied (eg Bergeron \& Leggett 2002). The largest LF measurement
currently published is that of Harris et al. (2006), using a sample of 6000 WDs selected from a combination of the SDSS 
and USNO-B. The presence of only 4 stars beyond the turning point precludes any accurate constraint on the LF at these 
magnitudes.

We are working on a program to measure the WDLF by exploiting data from the all-sky SuperCOSMOS Sky Survey (SSS). Sensitive
proper motion detection combined with accurate photometry over wide angles enables large samples of WDs to be 
catalogued using reduced proper motion methods. In section 2 we outline the SSS, and go on to descibe
our survey procedures in sections 3, 4 and 5. We present our LF in section 6,
and discuss the limitations and possible future refinements in section 7.

\section{The Data}
\label{data}
The SuperCOSMOS Sky Survey was compiled by digitizing several generations of photographic Schmidt plate surveys. The 
photographic source material includes the POSS-I and POSS-II surveys in the northern hemisphere, and a combination of
ESO and SERC surveys in the south. Over 1700 fields are required to cover the entire sky, with each field being observed
in four photographic bands $B_J$$R_{59F}$$I_N$ and $R_{103aE/63F}$ in the north/south respectively. 
%The observations span a baseline of up to fifty years, allowing for excellent proper motion detection with 
%uncertainties sometimes as low as $0.\!\!^{\prime\prime}004$~yr$^{-1}$. The plate depth varies from field to field, 
%with average magnitude limits of $B_J\sim22.6$, $R_{59F}\sim21.1$, $R_{103aE/63F}\sim21.0$ and $I_N\sim19.8$. 
Only $B_J$$R_{59F}$ and $I_N$ are normally used for 
photometry, $R_{103aE/63F}$ being generally of poorer quality and very similar in response to $R_{59F}$. $R_{103aE/63F}$ is 
%the band used in the POSS-I E and ESO-R surveys taken from 1949 to 1958 and from 1978 to 1990 respectively and is 
used to 
provide an early epoch for improved astrometric solutions in each field. 
%Note that in photographic nomenclature, $B_J$ is historically referred to simply as $J$, and $R$ sometimes as $E$. 
The subscripts refer to particular filter/emulsion
combinations, and will be dropped throughout the rest of this paper. We will sometimes refer to $R_{103aE/63F}$ and $R_{59F}$ 
as first- and second-epoch $R$ respectively.
Due to reasons outlined in section \ref{ProperMotionErrors}, we restrict our analysis to objects detected at all
four epochs.
The survey data is housed in relational databases available 
online at \verb"http://surveys.roe.ac.uk/ssa/". An introduction to the SSS and further technical details
can be found in Hambly et al. 2001a.
%
%\subsection{Survey field areas}
%Survey field solid angles are not given explicitly in the SSS literature, and have had to be calculated for this study. 
%Neighbouring field centres are offset by approximately $5^\circ$, resulting in $\sim0.5^\circ$ overlap between fields given
%the nominal $6^\circ$ field of view of Schmidt plates. The final merged source catalogue is seamless, with objects 
%observed multiple times in the plate overlap regions being assigned to fields whose centre they lie closest to along a 
%great circle. We measure the solid angle available to each field by simulating the system of field centres and breaking 
%down the sky into square arcminute sized regions, which are then assigned to their nearest field centre. 
%
%\subsection{Sky coverage}
We avoid the Galactic plane by $10^\circ$ and the centre by $20^\circ$. These regions are heavily affected by blending and 
poor photometric calibration due to differential extinction within the field. We also reject six fields centred
on the Magellanic clouds for the same reasons.
%We also reject fields 29, 51, 56, 57, 85 and 86. These are centred on the cores of the Magellanic clouds and suffer the
%same problems mentioned above.
This gives a total sky coverage of 10.2 sr, or about 80\% of the sky. Taking into account the regions lost
around bright stars due to diffraction spikes and scattered light haloes, the total survey area is 9.4 sr,
or just over 75\% of the sky.

%We also take account of the regions lost around bright stars due to diffraction spikes and halos. 
%This 'drilling fraction' is usually of the order of a few percent, but can reach as high as 90\% 
%in some southern hemisphere fields.

\section{Sample Selection}
\label{sample}
\subsection{Image morphology criteria}

The SuperCOSMOS Image Analysis Mode software measures 32 parameters for each object detection, some of which we
restrict in order to ensure our final WD catalogue only contains objects with reliable, high quality images. We
only accept objects with the \textit{deblending flag} set to zero, meaning no image fragmentation has been detected.
The \textit{quality number} is an integer indicating situations encountered during image analysis, such as proximity
to a bright star or plate boundary, and is set such that increasingly severe circumstances are assigned higher values.
We restrict this parameter to values less than 128. The \textit{profile classification statistic} $\eta$ gives a 
magnitude-independent measure of the `stellarity' of an image, by 
quantifying the deviation of the radial profile slope from that of the mean stellar template. It is scaled to a zero 
mean, unit variance Gaussian distribution. Objects with $|\eta| > 4\sigma$ are rejected.

%\subsubsection{Ellipticities of stellar images}
%
Previous studies utilising digitized Schmidt plate data have placed cuts on the ellipticities of images,
in order to limit contamination by faint galaxies and noise. However, we have found that the ellipticities measured
by SuperCOSMOS are extremely noisy at intermediate to faint magnitudes, and that any intuitively sensible cut will result in a 
seriously 
incomplete sample of stars. 
%On closer inspection, many real faint stellar images have been fitted with ellipses of eccentricity up to $\sim0.7$. 
This may be due in part to the eight-fold nearest neighbour 
pixel connectivity employed by the SuperCOSMOS image analysis software. We have decided to ignore this parameter.

\subsection{Contamination in north due to repeated photographic copying}

The northern half of the SSS data was digitized from photographic copies of POSS-I and POSS-II survey originals. 
Unfortunately, this seems to have degraded the quality of the data, resulting in erroneous astrometric fits
for a fraction of objects. This may be due to the photographic copying process rendering blended objects harder to 
detect. Certainly, the distribution of the astrometric residuals for affected objects deviates significantly from 
the expected $\chi^2$ form, indicating that something is going wrong with the centroiding.
%Unfortunately, the optical copying process seems to have suffered from a slight abberation, leading to an unexpected source of 
%contamination in the final digitized data. For the small fraction of affected stars, the abberation
%appears to throw off the image centroid as measured by SuperCOSMOS. This results in an erroneous astrometric fit, and 
%an unreliable proper motion. 
This problem first became obvious when we started to analyse the reduced proper
motion (RPM) diagram for the northern hemisphere (see section \ref{methods}), and noticed an extended population of 
objects at 
high values of RPM, in approximately the same region as cool WDs. This is what is expected if the proper motions of 
objects no longer correlate with distance - essentially, stars with bad proper motions are 
scattered from their usual locii and any discrimination between distinct Galactic and stellar components is lost. The 
main consequence for this work is that the highly numerous disk main sequence stars tend to scatter into the WD region,
completely swamping the cool WD locus.
Unfortunately, this contamination cannot adequately be removed by simply applying judicious selection cuts on the image 
parameters, and we have been forced to devise a new statistic to identify these objects.
This statistic is calculated using the known distributions of $\chi^2$ (astrometric fit residuals) and $\eta$ for both 
reliable and unreliable objects. The main result is that bad objects are successfully removed by placing a selection cut on this new
statistic, and that a certain fraction of good objects are also removed. This fraction can be precisely measured
and corrected for in later analysis, given that this statistic is independent of magnitude, colour and proper motion.

\section{Survey Limits}
\label{limits}
%Precisely defined and well characterised selection limits are important in any survey of this type. We wish to 
%maximise
%the accessible survey volume by extending the limits as far as possible without compromising the reliability of
%our final stellar catalogue - or at least, not in a way that cannot be measured and corrected for.

\subsection{Proper Motion}

The upper limit on proper motions is set to $0.\!\!^{\prime\prime}18$ yr$^{-1}$ for the entire survey. This arises from the mean epoch spread
and reliable pairing radius between consecutive observations in each field. This limit is not so important for 
producing large stellar catalogues, due to the lower probability of stars having such motions. Note, however, that 
nearby ultracool WDs and halo stars are expected to show larger motions. 
%and we plan to extend these limits in future work (see section \ref{FutureWork}). 
To produce a large sample of stars, we reduce the lower proper motion
limit as far as possible while maintaining a 5$\sigma$ proper motion detection threshold on all objects. This level has 
been chosen to limit observational scatter in the RPM diagram. Our lower limit
is thus dictated by the proper motion errors, which in general differ from field to field due to varying time baselines 
and plate quality, and within individual fields are a function of apparent magnitude, due mostly to centroiding 
precision as detailed in section 2.2.5 of Hambly et al 2001c. 
It is natural therefore to allow our lower proper motion limit to vary from field to field, and to make it
a function of apparent magnitude within each field. In general, there is no analytic form for the
distribution of proper motion errors with apparent magnitude, and a numerical solution has to be found.

\subsubsection{Proper motion limit as a function of apparent magnitude}
\label{ProperMotionErrors}

For every survey field, we plot the proper motion uncertainty $\sigma_{\mu}$ against $B$ for all objects.
Many fields share the same general form for this distribution, in particular a minimum of $\sigma_{\mu}\sim5$ mas yr$^{-1}$ at around $B\sim17$, rising to considerably 
larger errors at brighter and fainter magnitudes. Objects detected on only three plates show a similar distribution
offset to larger $\sigma_{\mu}$ by up to $0.\!\!^{\prime\prime}04$ yr$^{-1}$. Rather than attempting to define a 
separate lower proper motion limit
for these objects, we simply discard them. We wish to find the upper boundary of this distribution,
denoted $\sigma_{\mu}^{\textrm{{\scriptsize upper}}}(B)$. We may then define our $5\sigma$-threshold lower proper 
motion limit at every magnitude by simply multiplying this function by 5.
$\sigma_{\mu}^{\textrm{{\scriptsize upper}}}(B)$ is estimated by first binning objects according to magnitude, 
interactively so that each bin contains 100 points, which allows reasonable estimation of $\sigma_{\mu}^{\textrm{{\scriptsize upper}}}(B)$ at the extremes where objects are of low density. This results in $N$ bins, where $N = N_{\textrm{{\small objects}}}/100$
and is typically of the order of 100.  Each of the $N$ points $\sigma_{\mu}^{\textrm{{\scriptsize upper}}}(B_N)$ is then 
defined by the largest $\sigma_{\mu}$ at the mean $B$ value in each bin, after rejecting the top 3\% of points as
outliers. The set of points $\sigma_{\mu}^{\textrm{{\scriptsize upper}}}(B_N)$ shows considerable noise on small scales, and is 
processed through one stage of smoothing to remove this. A Savitzky-Golay filtering technique is used (Savitzky \& 
Golay 1964), generalized to allow for non-regular data.
This results in a final set of $N$ points $\sigma_{\mu}^{\textrm{{\scriptsize upper}}}(B_N)$, which are interpolated linearly
to define $\sigma_{\mu}^{\textrm{{\scriptsize upper}}}(B)$ at all magnitudes. Figure 1 shows an example of the proper
motion error distribution and $\sigma_{\mu}^{\textrm{{\scriptsize upper}}}(B)$ for one field.

%This method undoubtedly ends up rejecting many objects with good 5$\sigma$ detections but with proper motions below the
%lower limit. However, there is no way to include these without seriously compromising the proper motion
%completeness.

%%%%%%%%%%%%%%%%%%%%%%%%%%%%%%%%%%%%%%%%%%%%%%%%%%%%
%
% Figures
%
% 1) Proper motion error distribution + fitted upper boundary
%
% 2) Magnitude completeness histograms
%

%\begin{figure}[h]
\hspace{-1cm}
\begin{minipage}{8cm}
\includegraphics[angle=-90,width=8cm]{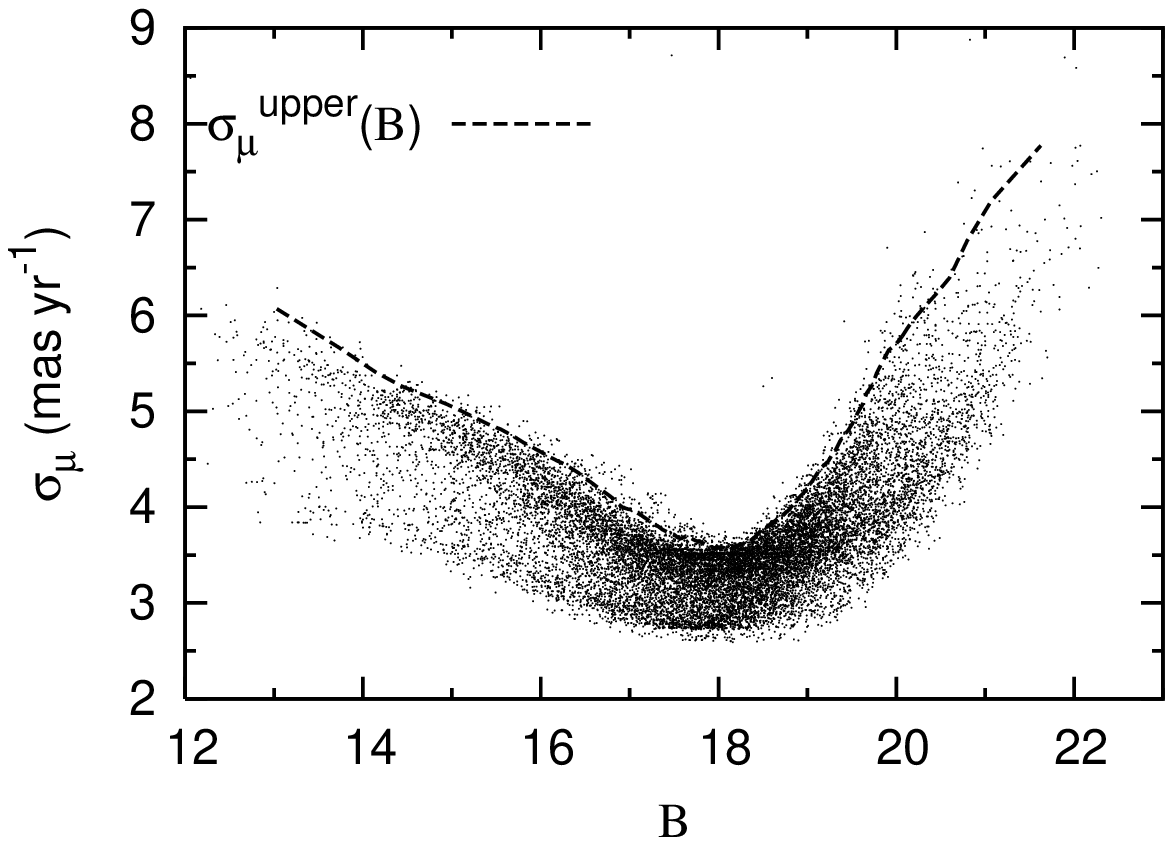}
\label{muerrors}
\hspace{1cm}
\begin{minipage}{6.5cm}
{\bf Figure 1: } {\footnotesize Proper motion error distribution for northern field 12, with the upper boundary set by $\sigma_{\mu}^{\textrm{{\scriptsize upper}}}(B)$}.
\end{minipage}
\end{minipage}
\begin{minipage}{8cm}
\includegraphics[width=8cm]{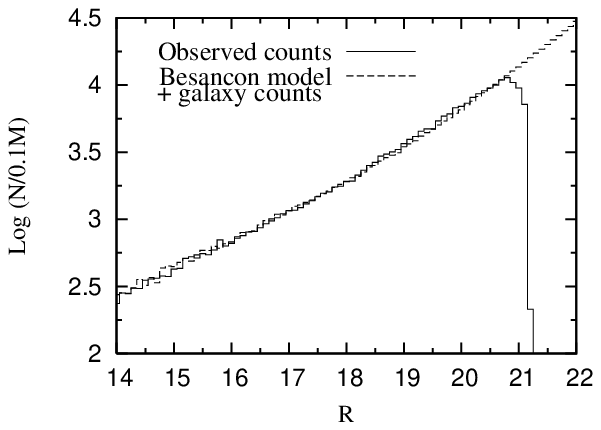}
\label{maghist}
\hspace{1cm}
\begin{minipage}{6.5cm}
{\bf Figure 2: } {\footnotesize Observed and theoretical number count histograms for SERC-R survey field 149.}
\end{minipage}
\end{minipage} 
%\end{figure}

%%%%%%%%%%%%%%%%%%%%%%%%%%%%%%%%%%%%%%%%%%%%%%%%%%%%

\subsection{Magnitude}
The magnitude completeness of catalogues derived from Schmidt plates has been analysed extensively in the literature.
We follow a similar approach to Tinney, Reid and Mould (1993), except that with nearly 8000 Schmidt plates in ~1700 
fields it is unfeasible to model the differential stellar and galactic number counts in each. We proceed by assuming
that within each of the eight photographic surveys compiled into the SSS the completeness of each plate is an
identical function of the plate detection limit. This allows for the considerable variation in depth of plates within 
each survey. We estimate the {\it completeness function} for each survey, using a subset of five fields chosen at
high Galactic latitude, in order to reduce the effects of image crowding and interstellar extinction. The
completeness function is defined to be the fraction of objects recovered as a function of apparent 
magnitude, and relies on an estimate of the intrinsic stellar and galactic number counts in each field.
Figure 2 shows the typical number count histograms for observed and modelled data in one field.

\subsubsection{Theoretical source counts}

Intrinsic stellar 
number counts are modelled using the Besancon Galaxy model (Robin et al. 2003). 
% one function of which is to provide number counts in a given survey field by integrating a range of simulated 
%Galactic stellar components along the line of sight. 
These will generally be different in each field due to different lines of sight through the Galaxy.
%\subsubsection{Theoretical galaxy counts}
Conversely, galaxy number counts are assumed to be isotropic, with the large ($\sim6^\circ\times6^\circ$) 
fields smoothing out any small scale structure.
Suitable empirical counts have been published by many authors (eg Jones et al. 2001), and in this study we use those
compiled by the Durham Cosmology Group\footnote{Available online at \texttt{http://star-www.dur.ac.uk/\~{}nm/pubhtml/counts/counts.html}}.
In log-number space the galaxy counts are well fitted by straight lines in each band over the range of interest.
% and are transformed to photographic $BRI$ bands then converted to $0.1M$ binning to match the 
%precision of the observed number count histograms.

%Galaxy number counts are modelled by the functions:
%\begin{eqnarray}
%\log(N_{\textrm{{\scriptsize Gal}}}/\textrm{{\small deg$^2$}}/\textrm{{\small 0.1 mag}}) &=& 0.4869B - 8.244 \\
%\log(N_{\textrm{{\scriptsize Gal}}}/\textrm{{\small deg$^2$}}/\textrm{{\small 0.1 mag}}) &=& 0.3823R - 5.43\\
%\log(N_{\textrm{{\scriptsize Gal}}}/\textrm{{\small deg$^2$}}/\textrm{{\small 0.1 mag}}) &=& 0.585I - 8.819  \qquad\;(I<18)\\
%\log(N_{\textrm{{\scriptsize Gal}}}/\textrm{{\small deg$^2$}}/\textrm{{\small 0.1 mag}}) &=& 0.3284I - 4.031 \qquad(I>18)
%\end{eqnarray}

\subsubsection{Completeness of digitized Schmidt plates as a function of plate depth}

%We first normalise the theoretical histograms to the observed histograms in each field at some suitable intermediate
%magnitude.
The completeness function in each field is given by $\cal C$$(m) = N_{\textrm{{\scriptsize obs}}}(m)/N_{\textrm
{{\scriptsize intrinsic}}}(m)$. For each survey we obtain five measures of $\cal C$$(m)$, which are individually 
shifted according to the plate depth then averaged
to obtain $\overline{\cal C}$$(m')$, where $m'$ is magnitudes above the plate limit. Uncertainties 
are assigned from the sample variance of the individual measures.
The resulting eight functions are used to define dynamic magnitude limits for the remaining fields in each survey.
One such function is presented in figure 3.
The offset from the plate limit chosen as the completeness limit ($m'_{lim}$) for each survey is given in table 1, along
with the corresponding mean magnitude limits for each band.

%The two $B$ band surveys appear to show significant incompleteness within $\sim2$ mags of the plate limits. However,
%this may be due in part to interstellar extinction affecting the observed data, which is not modelled in our 
%theoretical curves. As WDs would be affected only very slightly by extinction due to their relative proximity at
%a given apparent magnitude, we simply ignore this for now and set the magnitude limit by the final downturn in these
%functions.

%%%%%%%%%%%%%%%%%%%%%%%%%%%%%%%%%%%%%%%%%%%%%%%%%%%%
%
% Figures
%
% 3) Completeness functions
%
% Tables
%	
% 1) Table of survey depths + plate limit offsets
%

%\begin*{figure}[h]
\hspace{-1cm}
\begin{minipage}[H]{8cm}
\includegraphics[angle=-90,width=8cm]{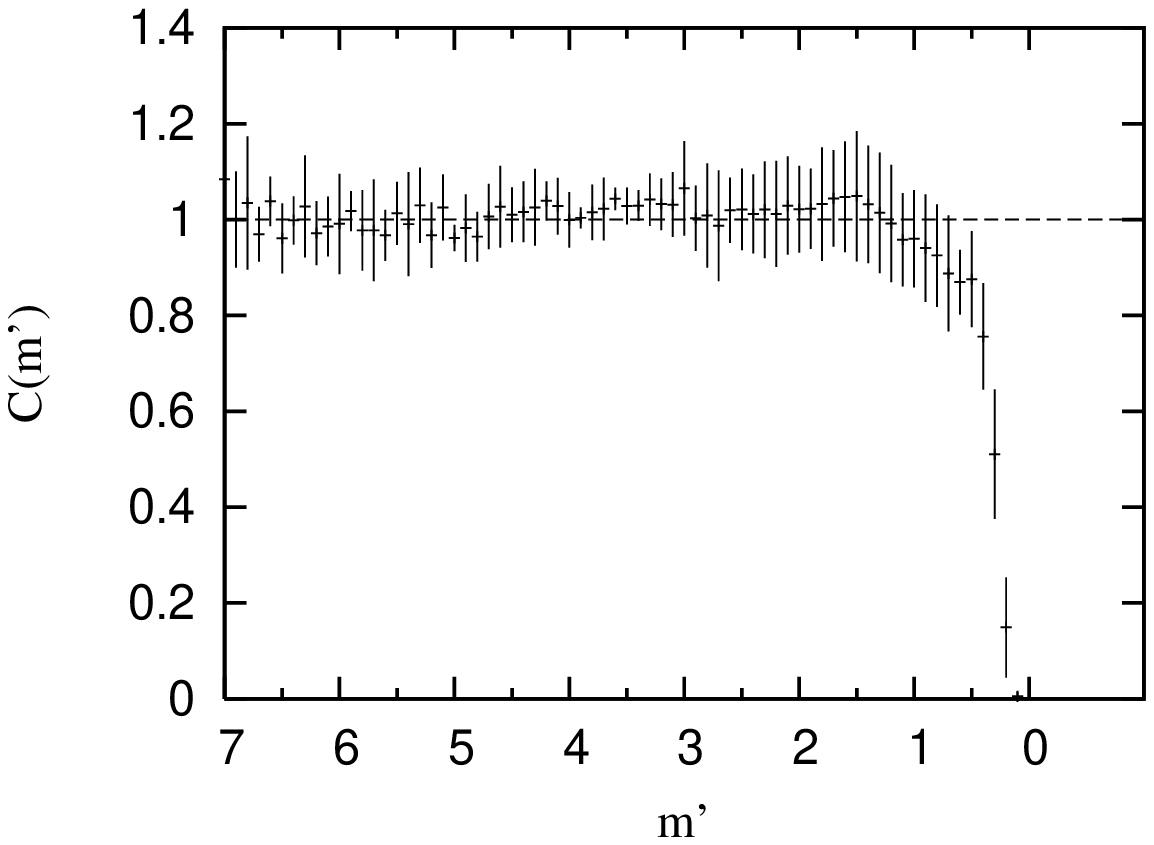}
\label{compfunc}
\hspace{1cm}
\begin{minipage}{6.5cm}
{\bf Figure 3:} {\footnotesize Mean completeness function for survey ESO-R. $m'$ refers to magnitudes above the plate limit; 
in this example $m'_{lim}=0.6$.}
\end{minipage}
\end{minipage}
\begin{minipage}{8cm}
\centering
\begin{tabular}{lcc}
\br
Survey&$m'_{lim}$&Mean depth\\
\mr
\multicolumn{3}{l}{Northern hemisphere:}\\
POSSI-E&0.8&20.2\\
POSSII-B&0.6&22.0\\
POSSII-R&0.7&20.5\\
POSSII-I&0.6&19.4\\
\multicolumn{3}{l}{Southern hemisphere:}\\
ESO-R&0.6&20.3\\
SERC-J&0.5&22.2\\
SERC-R&0.5&20.5\\
SERC-I&0.6&19.0\\
\br
\end{tabular}
\label{maglims}
\hspace{1cm}
\begin{minipage}{6.5cm}
{\bf Table 1:} {\footnotesize Completeness limit offsets and mean depth of surveys.}
\end{minipage}
\end{minipage}
%\end{figure}

%%%%%%%%%%%%%%%%%%%%%%%%%%%%%%%%%%%%%%%%%%%%%%%%%%%%

\section{Methods}
\label{methods}
\subsection{Reduced proper motion}

%The proper motions of nearby stars are correlated with distance, in the way that objects with larger
%angular velocities are likely to be closer than those with smaller velocities. For populations of objects
%with a finite distribution in tangential velocity such as disk stars, this assertion only holds in a statistical
%sense and cannot be used to obtain accurate distances, but is sufficient for separating stars of significantly
%different luminosity.

%The \textit{reduced proper motion} statistic $H_m = m + 5\log(\mu) + 5$ uses this correlation to estimate the absolute 
%magnitude of stars, 
The \textit{reduced proper motion} statistic $H_m = m + 5\log(\mu) + 5$ uses the correlation between proper motion
and distance to estimate the absolute magnitude of nearby stars,
and is commonly employed for identifying extremely sub-luminous WDs in proper motion surveys. 
The fact that $H_m$ can be expressed in quantities intrinsic to the star $H_m = M + 5\log(V_{tan}) - 
3.379$ suggests that we can use stellar and kinematic models to predict the range of $H_m$ for WDs of various 
colour and velocity. This allows us to perform rigorous selections on $H_m$ to produce well
defined catalogues of WD candidates.
This is done in the $H_m$-colour plane, where an appropriate colour-magnitude relation can be used to find the 
regions inhabited by WDs of various tangential velocity. Figure 4 shows the reduced proper motion diagram 
(RPMD) for 
our objects. This is topologically similar to the HR diagram. Cooling tracks for WDs with hydrogen and
helium atmospheres and a range of $V_{tan}$ are plotted. WDs of larger tangential
velocity are located below these lines, and we select objects from below the DA track as our WD candidates.

\subsubsection{Tangential velocity cut}

Cool WDs of low velocity share identical values of $H_m$ with subdwarfs in the high velocity tail of the halo, 
so these two populations overlap to some extent in the RPMD. Contamination by subdwarfs can be reduced 
by applying a minimum tangential velocity threshold to the kinematic selection of WDs. This essentially selects WD candidates 
from regions more widely separated from the subdwarf locus. The fraction of WDs that fall below the chosen $V_{tan}$ 
threshold can be corrected for by including a `discovery fraction'. This is calculated by projecting the 
WD velocity ellipsoid onto the sky, and integrating the resulting $V_{tan}$ distribution over
the range of allowed velocities, according to the prescription of Murray (1983). As each line of sight sees a slightly different projection of the velocity
ellipsoid, this has to be done separately for each survey field. 
We assume the velocity ellipsoid for disk WDs matches that of local dM stars measured by Reid et al 1995,
and set $\sigma_\textrm{u,v,w}$ = (43,31,25) kms$^{-1}$ with an asymmetric drift of -22 kms$^{-1}$ relative
to the local standard of rest. The discovery fraction has a mean value of $0.85$ ($0.69$) ($0.52$) for $V_{tan}>20 (30) (40)$kms$^{-1}$.
\subsection{Photometric parallaxes}
\label{photoparallax}
%Photometric distances and temperatures are obtained by fitting the two SSS colours to the models
%of Fontaine, Brassard \& Bergeron (2001), provided in the SuperCOSMOS passbands by P. Bergeron (private communication) and interpolated
%at 10K intervals. 
Photometric distances and temperatures are obtained by fitting the
two SSS colours using the model atmospheres and cooling sequences
described in Bergeron et al. (2001) and references therein (see also
\texttt{http://www.astro.umontreal.ca/\~{}bergeron/CoolingModels}), provided in
the SuperCOSMOS passbands and interpolated at 10K intervals.

The appropriate two-colour diagram and model tracks are presented in figure 6. We use a simple variance-weighted least-squares fitting procedure, and reject objects
with $\chi^2>5$. Such objects are mostly unresolved binaries, as detailed in section \ref{WDdM}.
With only one degree of freedom we cannot measure the gravity or atmosphere
%
%(two data points (B-R B-I), minus one free parameter (T_{eff}) )
%
type of each star, so we assume $\log(g)=8$ for
all stars and fit both hydrogen (DA) and helium (DB) atmospheres. Over- and under-massive WDs exist in roughly equal 
proportion, and fitting them
with normal-mass atmospheres has opposite effects and so should (roughly) cancel out. We make no
correction for reddening, as we expect the effect to be very small due to the proximity of our stars.
%%%%%%%%%%%%%%%%%%%%%%%%%%%%%%%%%%%%%%%%%%%%%%%%%%%%
%
% Figures
%
% 4) Reduced proper motion diagram
%
% 5) Col-mag relations
%
%\begin{figure}[h]
\begin{minipage}{7cm}
\includegraphics[width=7cm,height=8cm]{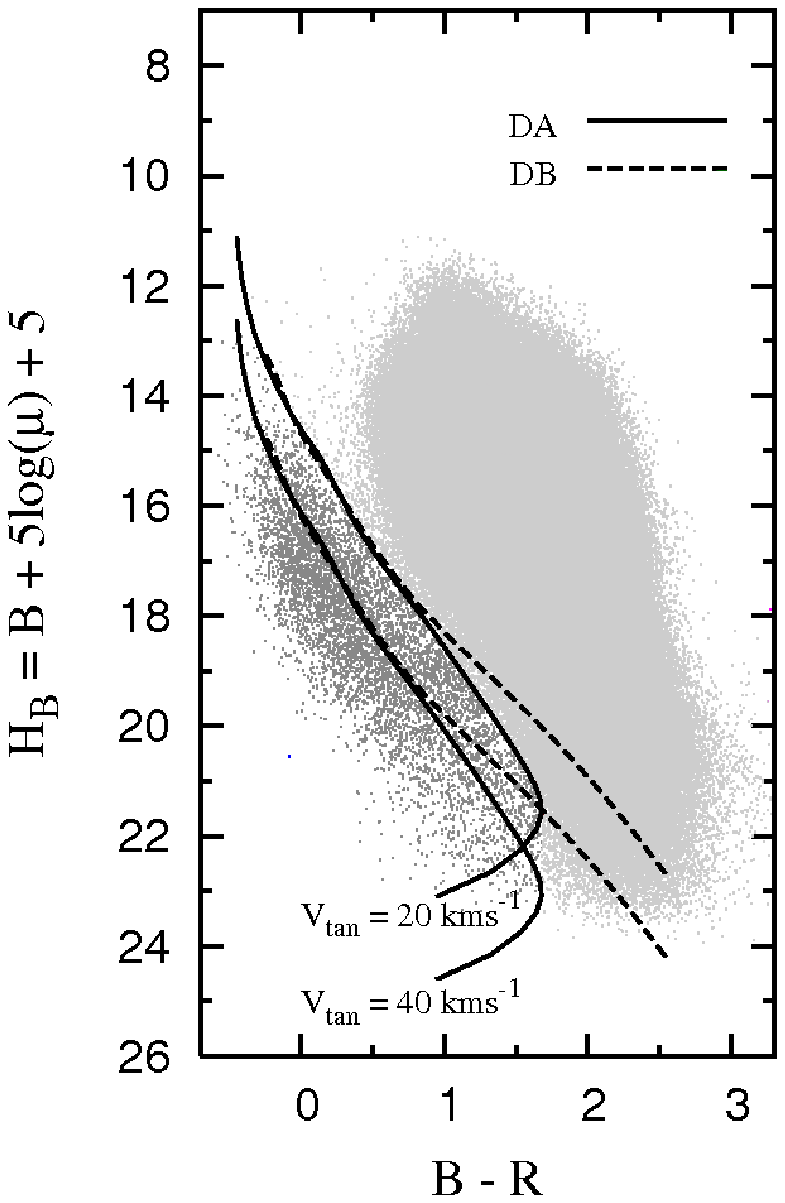}
\label{RPMD}
%\hspace{0.5cm}
\begin{minipage}{6.5cm}
{\bf Figure 4: } {\footnotesize RPMD for stars in SSS. Light grey points are subdwarfs and disk main sequence stars, dark
grey points are stars selected as WD candidates.}
\end{minipage}
\end{minipage}
\begin{minipage}{7cm}
\includegraphics[width=7cm,height=8cm]{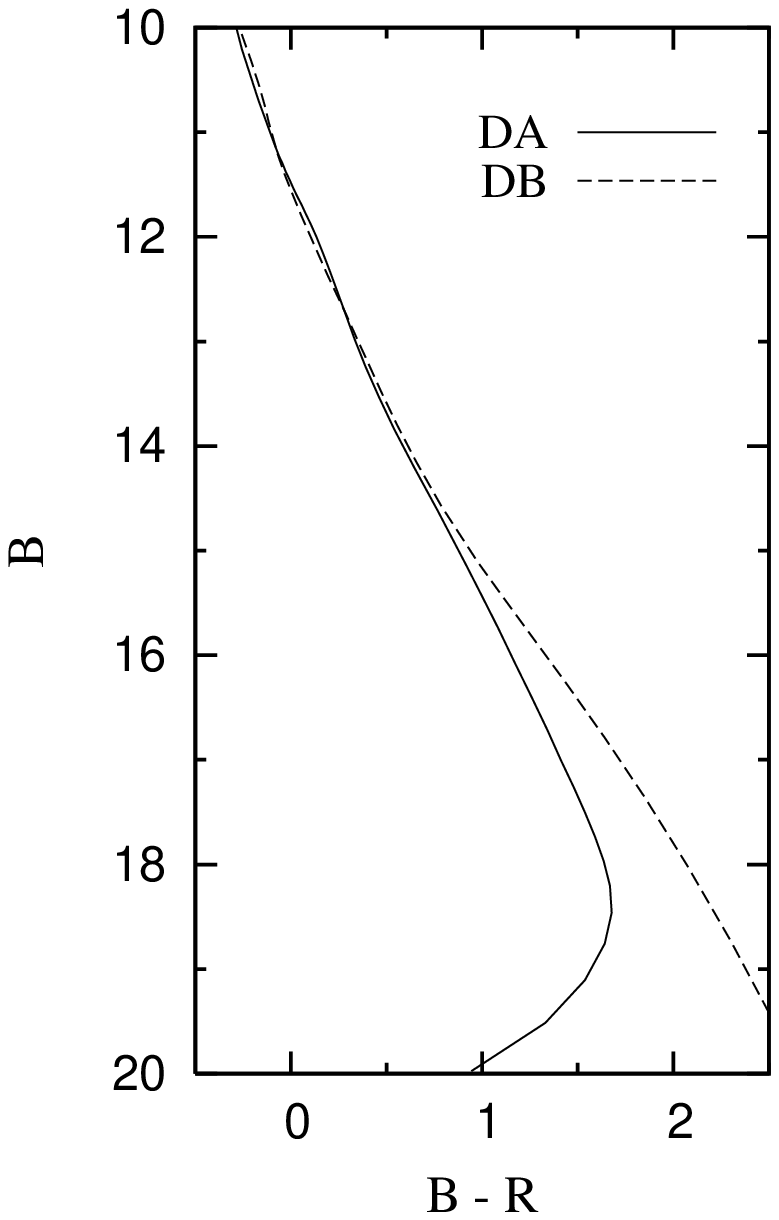}
\label{maghist}
\hspace{0.5cm}
\begin{minipage}{6.5cm}
{\bf Figure 5: } {\footnotesize Theoretical colour-magnitude relations for hydrogen-rich (DA) and helium-rich (DB) WDs, 
assuming a normal surface gravity of $\log(g)=8$.}
\end{minipage}
\end{minipage}
%\end{figure}
%%%%%%%%%%%%%%%%%%%%%%%%%%%%%%%%%%%%%%%%%%%%%%%%%%%%
\subsubsection{WD atmosphere types}
We allow WD candidates to contribute to the LF as both hydrogen and helium atmosphere WDs, with a weight for each type 
set according to the observed stellar abundances.
For WDs bluer than $B-R\sim1$, corresponding to $T_{eff}\gtrsim5250$K, the particular choice of atmosphere has 
little effect on the fitted luminosity, so we simply assume all stars are hydrogen types. At lower effective temperatures helium 
atmosphere stars are more
luminous, by around $1$ mag at $B-R\sim1.5$ (see figure 5), and we therefore expect helium WDs to be over-represented at the reddest 
colours. We shift the weight in favour of helium WDs in this range, and set it according to the fitted hydrogen 
atmosphere absolute magnitude. The values used are given in table 2.
\subsubsection{WD+dM binaries}
\label{WDdM}
WDs with unresolved cool companions will tend to scatter out of the WD locus in the colour-colour plane,
due to the companion star contributing flux to the I band. Harris et al. (2006) include these
objects in their survey by accepting only the bluest bands, however with only two colours we cannot do this
and so must reject these objects. This to some extent represents an incompleteness in our survey, as
around 10\% of WD candidates are rejected in this way.

%%%%%%%%%%%%%%%%%%%%%%%%%%%%%%%%%%%%%%%%%%%%%%%%%%%%
%
% Figures
%
% 6) Two colour plot
%
% Tables
%
% 2) DA/DB weights
%
%\begin{figure}[h]
\hspace{-1cm}
\begin{minipage}{8cm}
\includegraphics[angle=-90,width=8cm]{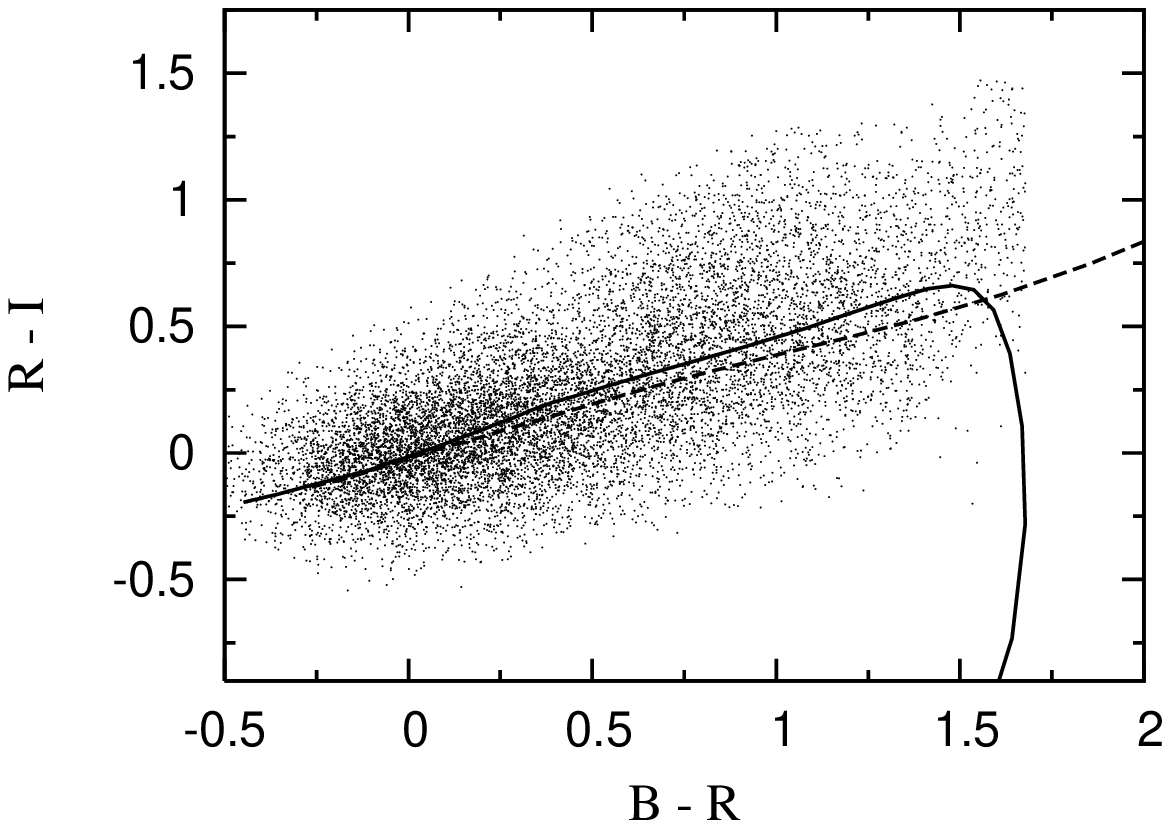}
\label{colplot}
\hspace{1cm}
\begin{minipage}{6cm}
{\bf Figure 6: } {\footnotesize Two-colour plot for WD candidates and synthetic WD models. The dashed line is for
helium-rich DB models, the solid line is for hydrogen-rich DAs.}
\end{minipage}
\end{minipage}
\begin{minipage}[c]{7cm}
\centering
\begin{tabular}{ccc}
\br
$M_{\textrm{bol, H}}$&${\cal W}_{\textrm{H}}$&${\cal W}_{\textrm{He}}$ \\
\mr

$<15.2$&$1.0$&$0$\\
$15.2\rightarrow15.5$&$0.5$&$0.5$\\
$15.5\rightarrow16.0$&$0.2$&$0.8$\\
$>16.0$&$0$&$1.0$\\

\br
\end{tabular}
\label{DADBweights}
\begin{minipage}[b]{6cm}
\vspace{0.5cm}
{\bf Table 2: } {\footnotesize Weights assigned to hydrogen and helium atmosphere WDs.}
\end{minipage}
\end{minipage} 
%\end{figure}
%%%%%%%%%%%%%%%%%%%%%%%%%%%%%%%%%%%%%%%%%%%%%%%%%%%%
%\subsection{Catalogue evaluation}
%Previously identified WDs, contamination from subdwarfs and M dwarfs, etc.
%SPectra for newly discovered objects.
\section{Luminosity function methods}
\label{wdlf}
\subsection{The generalized $V_{max}$ technique}

The spatial density of objects retrieved in a survey with both magnitude and proper motion
limits can be most readily obtained using Schmidt's $1/V_{max}$ method. This
involves summing the inverse of the \textit{maximum} volume in which each object could have been observed,
given it's absolute magnitude and tangential velocity, and the survey limits. This method has since been generalised
to non-uniformly distributed populations, in particular to allow for the exponential density profile of the
Galactic disk. $V_{max}$ is obtained by integrating the density law along the line of sight
to the star, over the distance range in which the star passes the survey selection criteria. 
%
%\[ V_{max} = \Omega \int_{\textrm{d$_{min}$}}^{\textrm{d$_{max}$}} r^2 \frac{\rho(r)}{\rho_{\odot}} dr \]
%
%where $\frac{\rho(r)}{\rho_{\odot}}$ is the density relative to the solar neighbourhood. 
%Poisson statistics are assumed
%for estimating errors, that is, each star carries an uncertainty equal to it's contribution, which are then summed
%in quadrature to obtain the error in each magnitude bin.
Errors are assigned to each luminosity bin by assuming Poisson statistics.

\subsubsection{$V_{max}$ with dynamic proper motion limits}
A problem arises when the proper motion limit is allowed to vary with apparent magnitude in a non-analytic way. The 
distances d$_{min}$ and d$_{max}$ at which the star's proper motion crosses the survey limits can
no longer be calculated analytically, and so this integral is evaluated in the following manner. The
limiting distances due to the magnitude limits are calculated as normal, then the integral is evaluated numerically
between these limits in steps of finite size. At each distance step, the star's proper motion is calculated and compared to 
the limits appropriate for the new apparent magnitude, and if the star is within the limits
then the volume associated with this step is added to the integral. The final stage is to select an appropriate 
step size for this calculation. We have chosen to use steps of constant \textit{generalized volume}, taking into
account the disk density profile. This ensures that each step makes an identical contribution to the sum. Note that 
calculating the distance interval necessary to keep the generalized volume of each step constant is
also non-trivial, and leads ultimately to a transcendental equation that must be solved numerically.

\subsection{The White Dwarf luminosity function}

The LF derived for the 9480 WDs with $V_{tan}>30$kms$^{-1}$ is presented in figure \ref{LF}.
A scaleheight of 250pc has been assumed for the density profile, to allow comparison with other works.
The LF is remarkably featureless on the rising side, apart from a slight inflexion at $M_{bol}\sim12$. The
faint end shows a drop in density at $M_{bol}=15.75$, in agreement with that reported by other authors.
Integrating the LF gives a total WD number density in the solar neighbourhood of $2.3\times10^{-3}$pc$^{-3}$.
$\langle\frac{V}{V_{max}}\rangle = 0.505 \pm 0.003$ for this survey, somewhat larger than that for a complete
survey ($\langle\frac{V}{V_{max}}\rangle=0.5$) but it should be noted that the assumed scaleheight affects
this calculation, and increasing it to 350pc results in $\langle\frac{V}{V_{max}}\rangle = 0.498 \pm 0.003$.

%%%%%%%%%%%%%%%%%%%%%%%%%%
%
%  WDLF with Vtan > 30 kms^-1
%
%%%%%%%%%%%%%%%%%%%%%%%%%%
\setcounter{figure}{6}
\begin{figure}[h]
\centering
\includegraphics[angle=-90,width=9cm]{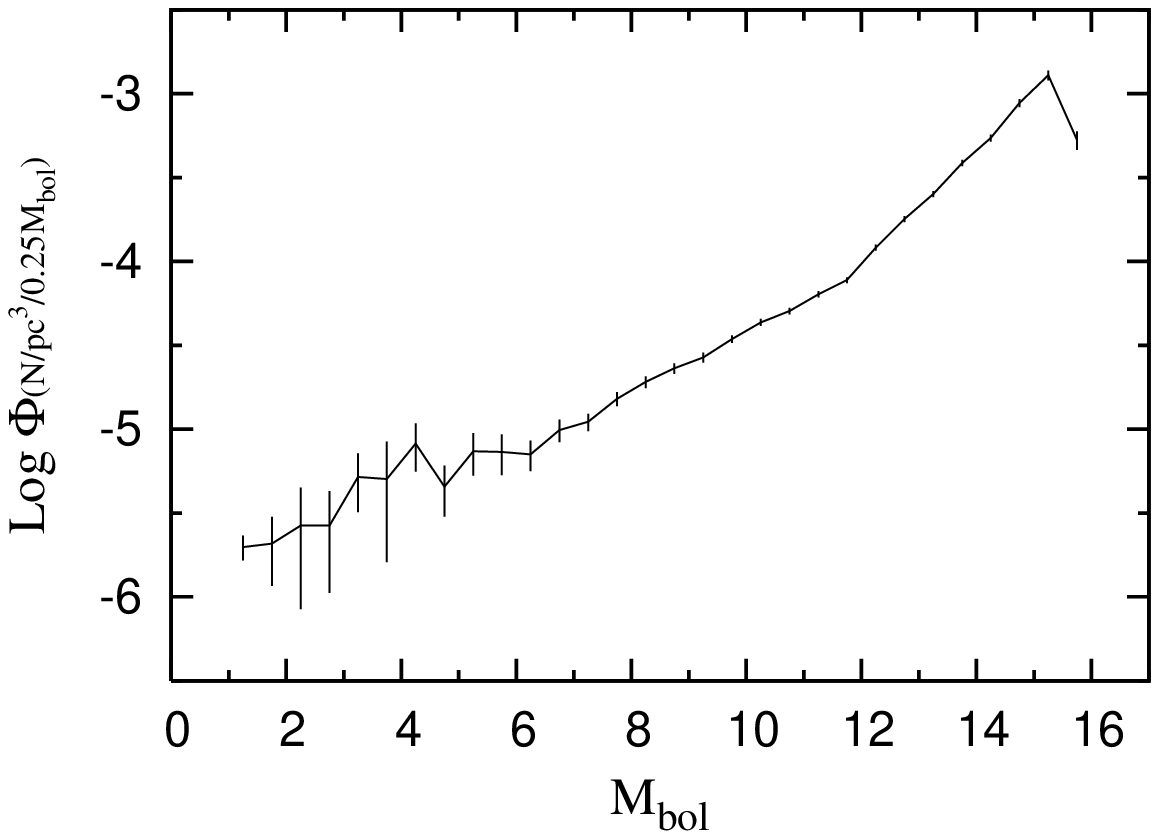}
\hspace{2pc}
\caption{LF for WDs with $V_{tan}>30$kms$^{-1}$.}
\label{LF}
\end{figure}

\section{Future Work}
 
\label{FutureWork}

Older stars are expected to have larger scaleheights due to kinematic heating, and $V_{max}$ will be 
underestimated for these objects by assuming a value of 250pc.
We intend to improve our LF by 
measuring the variation in scaleheight with $M_{bol}$ for our objects. Optical spectra have been obtained for a 
subsample of WD candidates, which we will use to assess the reliability of our method and contamination from non-
degenerate
stars. We also have a high proper motion catalogue of SSS objects with motions in the range 
$0.\!\!^{\prime\prime}18 < \mu < 10.\!\!^{\prime\prime}0$ yr$^{-1}$, which contains most of the interesting
ultracool WDs featured in the literature. Including this will push our disk LF to fainter magnitudes, and will also 
possibly provide enough halo WD candidates to measure a LF for this component.

\section*{References}

\begin{thereferences}

\item Bergeron P and Leggett S K 2002 {\it ApJ} {\bf580}:1070-1076

%\item Fontaine G, Brassard P and Bergeron P 2001 {\it PASP} {\bf113}:409-435
\item Bergeron P, Leggett S K and Ruiz MT 2001 {\it ApJ Suppl.} {\bf133}:413-449

\item Hambly N C {\it et al} 2001a {\it MNRAS} {\bf326}:1279-1294

\item Hambly N C, Davenhall A C, Irwin M J and MacGillivray H T 2001c {\it MNRAS} {\bf326}:1315-1327

\item Hansen, B {\it et al} 2007 {\it ApJ} {\bf671}:380-401

\item Harris H C {\it et al} 2006 {\it AJ} {\bf131}:571-581

\item von Hippel T and Gilmore G 2000 {\it AJ} {\bf120}:1384-1395

\item Jeffery E, von Hippel T, Jefferys W, Winget D, Stein N and DeGennaro S 2007 {\it ApJ} {\bf658}:391-395

\item Jones L R, Fong R, Shanks T, Ellis R S and Peterson B A 1991 {\it MNRAS} {\bf249}:481-497

\item Leggett S K, Ruiz M T and Bergeron P 1998 {\it ApJ} {\bf497}:294-302

\item Monteiro H, Jao W C, Henry T, Subasavage J and Beaulieu T 2006 {\it ApJ} {\bf638}:446-453

\item Murray C A 1983 {\it Vectorial Astrometry} 284-287

\item Noh H and Scalo J 1990 {\it ApJ} {\bf352}:605-614

\item Oswalt T, Smith J, Wood M and Hintzen P 1995 {\it Nat.} {\bf382}:692-694

\item Reid I N, Hawley S L and Gizis J E 1995 {\it AJ} {\bf110}:1838-1859

\item Robin A C, Reyl\'e C, Derri\'ere S and Picaud S 2003 {\it A \& A} {\bf409}:523-540

\item Savitzky A and Golay M 1964 {\it Analytical Chemistry} {\bf36}:1627-1639

\item Tinney C G, Reid I N and Mould J R 1993 {\it ApJ} {\bf414}:254-278

\item Winget D, Hansen C, Liebert J, Van Horn H, Fontaine G, Nather R, Kepler S and Lamb D 1987 {\it ApJ} {\bf315}:L77-L81

\end{thereferences}

\end{document}